\documentclass[twocolumn]{aastex63}

\usepackage[utf8]{inputenc}
\usepackage{hyperref}
\usepackage{amsmath}

\usepackage{silence}
\usepackage{graphicx}

\usepackage{float}
\usepackage{color,soul}
\usepackage{bm}

\usepackage{wasysym}
\usepackage{natbib}
\usepackage{makecell}

\usepackage{multirow}

\usepackage{xspace}

\newcommand\PaG{Pa$\gamma$\xspace}
\newcommand\PaB{Pa$\beta$\xspace}
\newcommand\BrG{Br$\gamma$\xspace}

\newcommand\Mdotpla{\ensuremath{\dot{M}_{\mathrm{pla}}}\xspace}
\newcommand\Lacc{\ensuremath{L_{\mathrm{acc}}}\xspace}
\newcommand\Fline{\ensuremath{F_{\mathrm{line}}}\xspace}
\newcommand\Lline{\ensuremath{L_{\mathrm{line}}}\xspace}
\newcommand\Lsun{\ensuremath{L_{\odot}}\xspace}
\newcommand\Mdot{\ensuremath{\dot{M}}\xspace}

\shorttitle{NIR Accretion in Delorme 1 (AB)b}
\shortauthors{Betti et al.}

\begin{document}

\title{Near-infrared Accretion Signatures from the Circumbinary Planetary Mass Companion Delorme~1~(AB)b
\footnote{Based on observations obtained at the Southern Astrophysical Research (SOAR) telescope, which is a joint project of the Minist\'{e}rio da Ci\^{e}ncia, Tecnologia e Inova\c{c}\~{o}es (MCTI/LNA) do Brasil, the US National Science Foundation’s NOIRLab, the University of North Carolina at Chapel Hill (UNC), and Michigan State University (MSU).} }

\author[0000-0002-8667-6428]{S. K. Betti}
\altaffiliation{Visiting astronomer, Cerro Tololo Inter-American Observatory at NSF’s NOIRLab, which is managed by the Association of Universities for Research in Astronomy (AURA) under a cooperative agreement with the National Science Foundation. }
\affiliation{Department of Astronomy, University of Massachusetts, Amherst, MA 01003, USA}

\author[0000-0002-7821-0695]{K. B. Follette}
\affiliation{Department of Physics and Astronomy, Amherst College, Amherst, MA 01003, USA}

\author[0000-0002-4479-8291]{K. Ward-Duong}
\affiliation{Department of Astronomy, Smith College, Northampton, MA, 01063, USA}

\author[0000-0003-0568-9225]{Y. Aoyama}
\affiliation{Institute for Advanced Study, Tsinghua University, Beijing 100084, PR China}
\affiliation{Department of Astronomy, Tsinghua University, Beijing 100084, PR China}

\author[0000-0002-2919-7500]{G.-D. Marleau}
\affiliation{Institut f\"ur Astronomie und Astrophysik, Universit\"at T\"ubingen, Auf der Morgenstelle 10, 72076 T\"ubingen, Germany}
\affiliation{Physikalisches Institut, Universit\"at Bern, Gesellschaftsstr.~6, 3012 Bern, Switzerland }
\affiliation{Max-Planck-Institut f\"ur Astronomie, K\"onigstuhl 17, 69117 Heidelberg, Germany}

\author{J. Bary}
\affiliation{Department of Physics and Astronomy, Colgate University, Hamilton, NY, 13346, USA}

\author[0000-0003-1639-510X]{C. Robinson}
\affiliation{Department of Physics and Astronomy, Amherst College, Amherst, MA 01003, USA}

\author[0000-0001-8345-593X]{M. Janson}
\affiliation{Institutionen f\"or astronomi,
AlbaNova universitetscentrum,
Stockholms universitet, 10691 Stockholm, Sweden}

\author[0000-0001-6396-8439]{W. Balmer}
\affiliation{The William H. Miller III Department of Physics \& Astronomy, Johns Hopkins University, Baltimore, MD 21218, USA}
\affiliation{Space Telescope Science Institute, 3700 San Martin Drive, Baltimore MD 21218, USA}

\author[0000-0003-4022-8598]{G. Chauvin}
\affiliation{Laboratoire Lagrange, Université Cote d’Azur, CNRS, Observatoire de la Cote d’Azur, 06304 Nice, France}

\author[0000-0002-6217-6867]{P. Palma-Bifani}
\affiliation{Laboratoire Lagrange, Université Cote d’Azur, CNRS, Observatoire de la Cote d’Azur, 06304 Nice, France}

\correspondingauthor{S. K. Betti}
\email{sbetti@umass.edu}

\begin{abstract}
Accretion signatures from bound brown dwarf and protoplanetary companions provide evidence for ongoing planet formation, and accreting substellar objects have enabled new avenues to study the astrophysical mechanisms controlling formation and accretion processes.  Delorme 1 (AB)b, a $\sim30$--45~Myr circumbinary planetary mass companion, was recently discovered to exhibit strong H$\alpha$ emission. This suggests ongoing accretion from a circumplanetary disk, somewhat surprising given canonical gas disk dispersal timescales of $5-10$ Myr. Here, we present the first NIR detection of accretion from the companion in Pa$\beta$, Pa$\gamma$, and Br$\gamma$ emission lines from SOAR/TripleSpec 4.1, confirming and further informing its accreting nature.  The companion shows strong line emission, with $\Lline \approx 1$--$6 \times 10^{-8}~\Lsun$ across lines and epochs, while the binary host system shows no NIR hydrogen line emission (\Lline $<0.32-11\times10^{-7}$ \Lsun).  Observed NIR hydrogen line ratios are more consistent with a planetary accretion shock than with local line excitation models commonly used to interpret stellar magnetospheric accretion.     
Using planetary accretion shock models, we derive mass accretion rate estimates of $\Mdotpla\sim3$-$4\times 10^{-8}\ M_\mathrm{J}$ yr$^{-1}$, somewhat higher than expected under the standard star formation paradigm. Delorme 1 (AB)b's high accretion rate is perhaps more consistent with formation via disk fragmentation.
Delorme 1 (AB)b is the first protoplanet candidate with clear (S/N$\sim$5) NIR hydrogen line emission. 
\end{abstract}

\section{Introduction} \label{sec:intro}

The theory of magnetospheric accretion, whereby infalling inner disk material flows along stellar magnetic field lines and forms a shock in a young star’s atmosphere, is well-established and consistent with a range of observations \citep[e.g.,][]{Koenigl1991}.  X-ray emission originating from the shock front is absorbed and re-radiated as excess optical/ultraviolet Balmer continuum \citep[e.g.][]{Hartmann2016, Valenti1993, Gullbring1998, Calvet1998}, while infalling gas exhibits line emission, including the Balmer, Paschen, and Brackett series hydrogen lines.

The same accretion process has been assumed to extend to substellar masses \citep[e.g.][]{Muzerolle2005, Alcala2017}, and accretion signatures from planetary mass companions (PMCs) have been interpreted under the stellar paradigm. Recent discoveries of H$\alpha$ accretion signatures in substellar companions---both brown dwarfs (BD) \citep[e.g.,~SR12c;][]{Santamaria-Miranda2018,Santamaria-Miranda2019} and protoplanet candidates \citep[e.g., PDS 70 b and c and LkCa 15 b,][]{Haffert2019, Wagner2018, Sallum2015}---have provided incontrovertible evidence of accretion onto secondary objects in young systems.   Combined with the first detections of circumplanetary disks \citep{Benisty2021}, these systems allow for direct study of planet formation processes.

Recently, \citet{Eriksson2020} discovered strong hydrogen (H$\alpha$, H$\beta$) and helium emission lines suggestive of ongoing accretion from the PMC 2MASS J01033563-5515561(AB)b, also known as Delorme 1 (AB)b.  Among the first imaged circumbinary PMCs, Delorme 1 (AB)b was discovered in $L'$ band by \citet{Delorme2013}  at 1\farcs77 (84 au) separation, with an estimated mass of 12--14 $M_\mathrm{Jup}$, placing it at the deuterium burning limit.  Its host, Delorme 1 AB, is an M5.5 binary \citep[separation of $\sim$0\farcs25 or 12 AU;][]{Delorme2013} at 47.2 $\pm$ 3 pc \citep{Riedel2014} in the Tucana-Horologium association \citep{Gange2015}, placing its age at $\sim$30--45~Myr.
While the system shows evidence of youth, including an overluminous central binary \citep{Riedel2014}, red $JHK_s$ colors \citep[similar to other young bound nonaccreting companions;][]{Riedel2014}, and low surface gravity \citep{Liu2016}, 
ongoing PMC accretion at 30--45~Myr is possible, as lower mass objects have been found to have long disk dispersal timescales \citep{Luhman2022}\footnote{\citet{Luhman2022} found that in the 15-21 Myr Lower Centaurus Crux and Upper Centaurus Lupus association, disk fractions increase with decreasing mass, from 0.7\% to 9\%, indicating lower-mass stars can retain disks far longer than originally estimated ($\sim$10 Myr).}.

In this letter, we present the first detection of near infrared (NIR) emission lines from Delorme 1 (AB)b, corroborating the claim of ongoing companion accretion, and confirming the lack of accretion in the host binary system.  This is the first accreting PMC with \PaB, \PaG, and Br$\gamma$ detections, and provides a critical benchmark for future NIR accretion studies of PMCs. NIR line ratios provide an important probe of the physical properties of the emitting region that can inform accretion paradigms.

\begin{figure*}[hpt!]
    \centering
    \includegraphics[width=\linewidth]{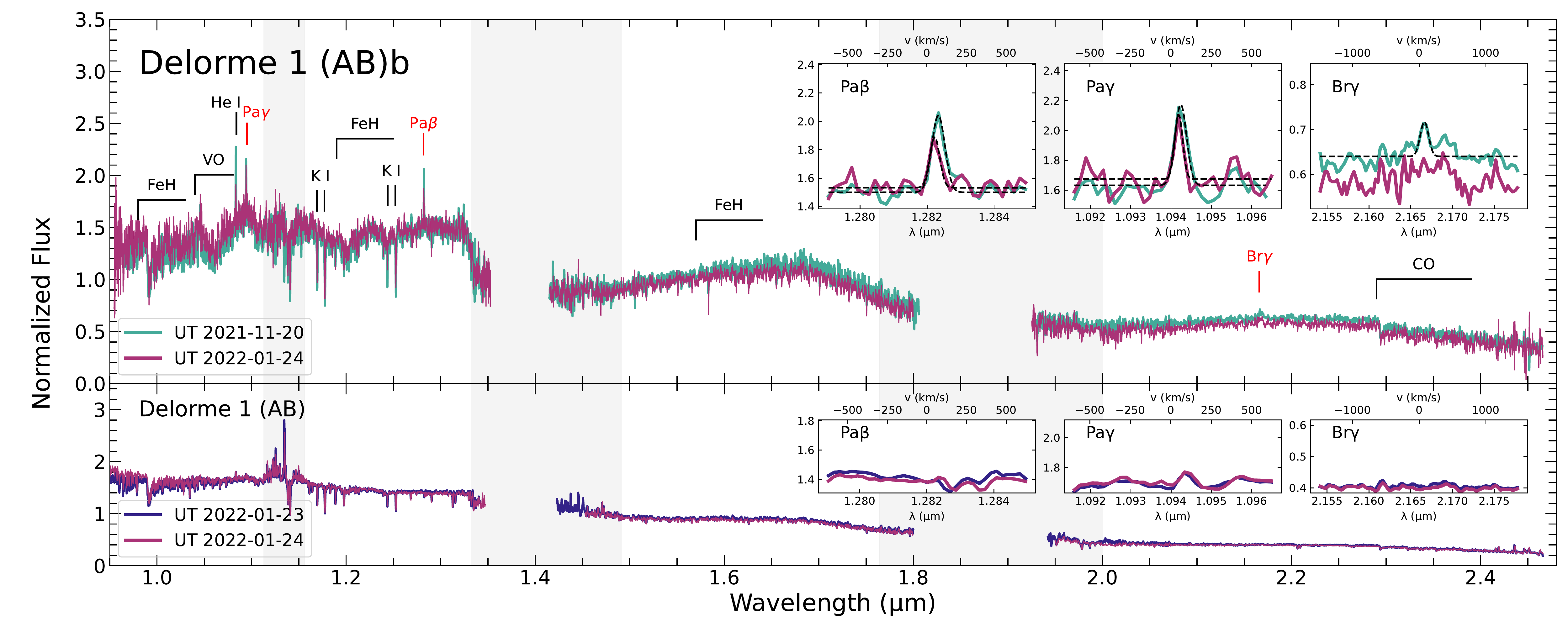}
    \caption{TripleSpec4.1 $JHK$-band spectra of Delorme 1 (AB)b (\textit{top;} green: UT 2021-11-20, magenta: UT 2022-01-24) and Delorme 1 AB (\textit{bottom;} blue: UT 2022-01-23, magenta: UT 2022-01-24)
    The NIR emission lines, Pa$\gamma$, Pa$\beta$, and Br$\gamma$, are highlighted with red labels. The lines are shown in greater detail in the inset images for the companion (\textit{top}, with best fit Gaussians shown by dashed black lines) and binary host (\textit{bottom}).  Other atomic and molecular features are labeled in black. Gray bands indicate regions of high atmospheric absorption.}
    \label{fig1}
\end{figure*}

\section{Observations and Reductions}

Delorme 1 (AB)b was observed with the TripleSpec 4.1 Near-IR spectrograph \citep{Schlawin2014} on the Southern Astrophysical Research (SOAR) Telescope during two observing runs in 2021-2022 (ID: 2021B-0311). TripleSpec 4.1 covers 0.8-2.47 $\mu$m at moderate resolution (R$\sim$3500) with a fixed 1\farcs1 $\times$ 28\arcsec slit.  Both observations were taken in good weather conditions, with seeing around 0\farcs95-1\farcs0, with the slit aligned to the parallactic angle.  Delorme 1 (AB)b was observed on 2021 November 20 (epoch 1) at an airmass of 1.2.  Sixteen 180~s exposures were taken in an ABBA cycle, for a total exposure time of 2880 s, yielding a final reduced spectrum with a mean SNR of $\sim90$ at $H$-band.  On 2022 January 24 (epoch 2), we observed Delorme 1 (AB)b at an airmass of 1.27 with an identical observational strategy and total integration time, with the reduced spectrum achieving a mean SNR of $\sim60$ at $H$-band. We observed the binary Delorme 1 AB on 2022 January 23 (airmass 1.34), and on 2022 January 24 (airmass 1.65). We took eight 30-s exposures in an ABBA cycle, for a total of 240~s each night, yielding average final spectrum SNRs of 270 and 300 at $H$-band.  As the seeing on January 23 was $\sim$1\farcs3, we were not able to sufficiently resolve the companion and did not attempt to observe it. 

Spectra were reduced using a TripleSpec 4.1 version of SpeXtool \citep{Cushing2004} following the standard procedure: subtraction of A and B frames for sky removal, order identification, spectral extraction, and wavelength calibration from arc lamps. The orders were merged and areas of significant atmospheric absorption removed.  A spectrophotometric standard (HIP 6364, A0V) was observed before and after Delorme 1 for both telluric correction and flux calibration, following \citet{Vacca2003} using the SpeXtool \texttt{xtellcor} software.  
Due to its close distance \citep[$47.2 \pm 3.1$~pc;][]{Riedel2014}, Delorme 1 resides in the Local Bubble \citep[area of low interstellar extinction;][]{Sfeir1999}; therefore, we assume zero reddening.  

\section{Results}
We detect strong \PaB, \PaG, and \BrG emission lines (Figure~\ref{fig1}) in Delorme 1 (AB)b in both epochs.   
Hydrogen emission lines are not detected in the host binary (see Table~\ref{tab:results} for line flux upper limits), 
providing strong confirmation they are unique to the companion.  

We compute equivalent widths (EW), fluxes (\Fline) and luminosities (\Lline) for each line and epoch (Table~\ref{tab:results}).
Line fluxes are computed by integrating under a best-fit Gaussian profile after subtracting a linear fit to the local continuum.  The uncertainty on the line is a function of the scatter in the continuum and the best-fit Gaussian given by
\begin{equation}
    \sigma = \sqrt{N_\mathrm{pix}} \times F_\mathrm{noise} \times \Delta\lambda, 
\end{equation}
where $N_\mathrm{pix}$ is the number of pixels within 3$\times$FWHM, $F_\mathrm{noise}$ is the rms of the local continuum, and $\Delta\lambda$ is the wavelength resolution per pixel at each line.  EWs are obtained from the ratio of line fluxes to the average local continuum level within a 50 \AA{} window on either side of the line.  We estimate EW uncertainties following \citet{Vollmann2006}.

We do not detect \BrG in epoch 2, potentially due to poorer seeing conditions. For non-detected lines, we calculate \Fline upper limits as $F_\mathrm{line}^\mathrm{upp} = 3\sigma$.   

\begin{deluxetable*}{@{\extracolsep{4pt}}cccccc@{}ccc}
\tablenum{1}
\tablecaption{Delorme 1 (AB)b Line Characteristics\label{tab:results}}
\tablewidth{0pt}
\tablehead{
\colhead{} & \colhead{} & \colhead{} & \colhead{} & \multicolumn{2}{c}{stellar scaling\tablenotemark{a}} & \multicolumn{2}{c}{planetary scaling\tablenotemark{b}} & \multicolumn{1}{||c}{} \\
\cline{5-6}
\cline{7-8}
\colhead{Line} & \colhead{EW} & \colhead{\Fline} & \colhead{\Lline} & \colhead{$\log(\Lacc)$} & \colhead{$\log(\Mdot)$} & \colhead{$\log(\Lacc)$} & \colhead{$\log(\Mdot)$} &  \multicolumn{1}{||c}{\multirow{-1.5}{*}{\parbox{2.5cm}{\centering Delorme 1 AB \Fline}}} \\
\colhead{}  & \colhead{(\AA)} & \colhead{(10$^{-16}$ erg/s/cm$^2$)} & \colhead{($10^{-8} \Lsun$)} & \colhead{($\Lsun$)} & \colhead{($ M_\mathrm{J}$ yr$^{-1}$)}& \colhead{($\Lsun$)} & \multicolumn{1}{c}{($ M_\mathrm{J}$ yr$^{-1}$)} & \multicolumn{1}{||c}{(10$^{-15}$ erg/s/cm$^2$)}
}
\startdata
\multicolumn{8}{c}{UT 2021-11-20} & \multicolumn{1}{||c}{}\\
\hline
\PaG	&	-1.95$\pm$0.74	&	6.82$\pm$1.33	&	4.75$\pm$1.11	&	-5.50$\pm$0.53	&	-8.82$\pm$0.53	&	-3.94$\pm$0.30	&	-7.27$\pm$0.30	&	\multicolumn{1}{||c}{$<$3.67}\\	
\PaB	&	-2.31$\pm$0.88	&	8.05$\pm$1.49	&	5.60$\pm$1.27	&	-4.92$\pm$0.62	&	-8.25$\pm$0.62	&	-4.02$\pm$0.30	&	-7.35$\pm$0.30	&	\multicolumn{1}{||c}{$<$3.71} \\	
\BrG	&	-2.08$\pm$1.11	&	1.64$\pm$0.56	&	1.14$\pm$0.42	&	-5.43$\pm$0.94	&	-8.75$\pm$0.96	&	-3.91$\pm$0.30	&	-7.24$\pm$0.30	&	\multicolumn{1}{||c}{$<$1.01}\\	
\hline
\multicolumn{8}{c}{UT 2022-01-24} & \multicolumn{1}{||c}{}\\
\hline																
\PaG	&	-1.24$\pm$0.47	&	2.94$\pm$0.77	&	2.05$\pm$0.61	&	-5.95$\pm$0.55	&	-9.28$\pm$0.56	&	-4.25$\pm$0.30	&	-7.58$\pm$0.30	&	\multicolumn{1}{||c}{$<$7.91} \\	
\PaB	&	-1.44$\pm$0.62	&	3.49$\pm$0.85	&	2.43$\pm$0.67	&	-5.31$\pm$0.64	&	-8.64$\pm$0.65	&	-4.33$\pm$0.30	&	-7.67$\pm$0.30	&	\multicolumn{1}{||c}{$<$6.56} \\	
\BrG	&	--	&	$<$0.74	&	$<$0.52	&	$<$-5.81	&	$<$-9.17	&	$<$-4.20	&	$<$-7.53	&	\multicolumn{1}{||c}{$<$1.86} \\	
\enddata
\tablenotetext{a}{\Lacc--\Lline scaling relation from \citet{Alcala2017}}
\tablenotetext{b}{\Lacc--\Lline scaling relation from \citet{Aoyama2021}}
\end{deluxetable*} 

During magnetospheric accretion, the infalling column of gas is heated to $\sim10^4$ K, producing broad emission lines \citep{Hartmann2016} such as \PaB, \PaG, and \BrG.  The gas travels at free-fall velocity, and heats to $10^6$ K when it shocks at the stellar photosphere, fully ionizing and preventing the formation of hydrogen line emission.  
In contrast, recent simulations of accreting PMCs \citep{Aoyama2018, Aoyama2020} suggest differences in the physical conditions of the shocked region. Due to smaller masses and lower surface gravities, accreting gas travels at lower free fall velocities, leading to a non-fully ionized post-shock region.  This results in shock-heated accreting gas capable of hydrogen line emission \citep{Aoyama2018}.  
In other words, the detections of Paschen and Brackett-series emission from accreting objects are an unambiguous sign of accretion; however, the dominant source of line emission may be either the infalling accretion column or the post-shock region.
Given this ambiguity, we estimate accretion rates for Delorme 1 (AB)b following both families of accretion models, and discuss the differences below.
The mass accretion rate is given by:
\begin{equation}
    \dot{M} = \left(1-\frac{R_\star}{R_\mathrm{in}}\right)^{-1} \frac{\Lacc R_\star}{GM_\star},
\end{equation}
where $R_\mathrm{in}$ is the inner disk radius, assumed to be 5 $R_\star$ \citep[e.g.,][]{Gullbring1998, Herczeg2008, Rigliaco2012, Alcala2017}, $R_\star$ is the radius of the accreting object, $M_\star$ is its mass, and \Lacc is the estimated accretion luminosity.

Total accretion luminosity has been found to strongly correlate with emission line luminosities in T Tauri stars \citep{Rigliaco2012, Alcala2014, Alcala2017} as
\begin{equation}
    \log (L_\mathrm{acc}/L_\odot) = a\times\log (L_\mathrm{line}/L_\odot) + b,
\end{equation}
where $a$ and $b$ are the fit coefficients for each line.
These relationship can be used to estimate $\dot{M}$ from a single accretion-tracing line.  However, \citet{Aoyama2020, Aoyama2021} argue that the \Lline--\Lacc relationships are not valid for planetary mass objects because of the different physical conditions of the emitting region.  \citet{Aoyama2021} derived new theoretical \Lacc--\Lline relationships expected for PMCs based on the \citet{Aoyama2018} shock models.  We refer to all accretion luminosities and mass accretion rates derived from \citet{Alcala2017} \Lacc--\Lline scaling relations as ``stellar" (e.g., ~$L_\mathrm{acc, ste}$/$\dot{M}_\mathrm{ste}$) and those derived from \citet{Aoyama2021} as ``planetary" (e.g., ~$L_\mathrm{acc, pla}$/$\Mdotpla$) for ease of distinguishing between the two.  

 \begin{figure}[tp]
     \hspace{-0.6cm}
     \includegraphics[width=0.5\textwidth]{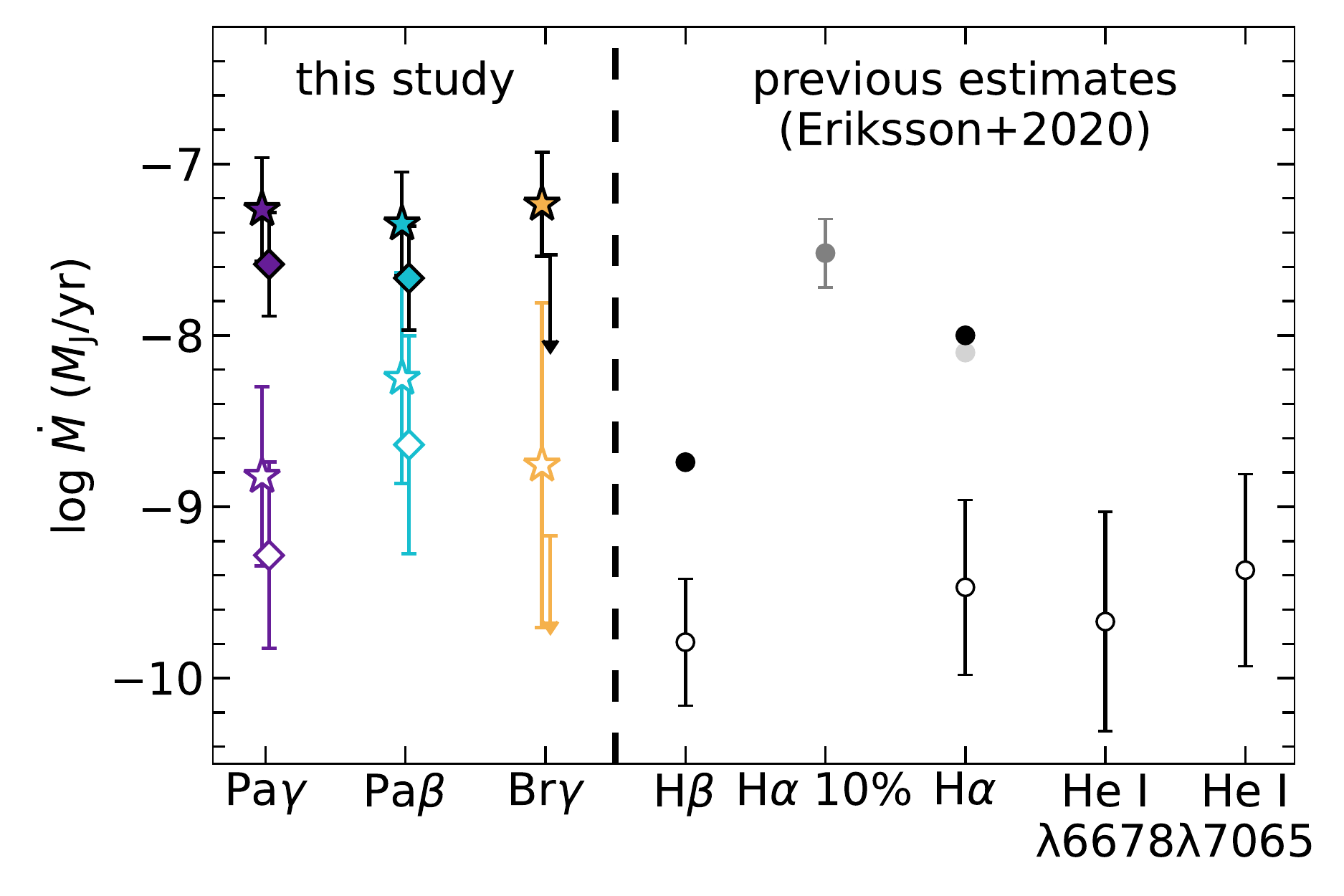}
     \caption{Accretion rates for individual emission lines. Circles indicate data from \citet{Eriksson2020}. Stars and diamonds represent our epochs 1 and 2, respectively.
     Accretion rates derived using both ``stellar" empirical scaling relations \citep[open symbols, ][]{Alcala2017} and ``planetary" accretion models \citep[filled symbols, ][]{Aoyama2021} are shown. \Mdot(H$\alpha$) is also estimated using the line luminosity model of \citet[][light gray]{Thanathibodee2019}.   
     \Mdot(H$\alpha$ 10\%) is estimated using a ``stellar" empirical relation \citep[dark gray, ][]{Natta2004}. } 
     \label{fig2}
 \end{figure}

Following \citet{Eriksson2020}, we assume a companion mass of $M_p = 0.012\ M_\odot$ and radius $R_p = 0.163\ R_\odot$.  We calculate $\Lacc$ following the \Lacc--\Lline scaling relations calibrated empirically for stars (\PaB: ($a,b$)=(1.06, 2.76), \PaG: ($a,b$)=(1.24, 3.58), \BrG: ($a,b$)=(1.19, 4.02)) by \citet{Alcala2017} and theoretically for PMCs (\PaB: ($a,b$)=(0.86, 2.21), \PaG: ($a,b$)=(0.85, 2.28), \BrG: ($a,b$)=(0.85, 2.84)) by \citet{Aoyama2021}. This allows us to directly compare our NIR-derived results to the accretion rates estimated by \citet{Eriksson2020}. 

Our \Mdot estimates are given in Table~\ref{tab:results} for both the ``stellar" and ``planetary" relations.  \Mdot estimates are relatively consistent among lines and epochs under each scaling relation; however, the \citet{Aoyama2021} models predict \Mdot's that are systematically higher by several orders of magnitude.

On average, using the stellar scaling relations of \citet{Alcala2017} we find a
$\log(\dot{M}_\mathrm{ste})$ of $-8.53 \pm 0.28 \ M_\mathrm{J}\ \mathrm{yr}^{-1}$ for epoch 1 and  $-8.85\pm0.28\ M_\mathrm{J}\ \mathrm{yr}^{-1}$ for epoch 2. Using the \citet{Aoyama2021} planetary shock-model relations, we find $\log(\Mdotpla) = -7.38\pm0.23\ M_\mathrm{J}\ \mathrm{yr}^{-1}$ and $-7.19\pm0.31\ M_\mathrm{J}\ \mathrm{yr}^{-1}$ for epochs 1 and 2, respectively. 

\begin{figure*}[htp!]
    \centering
    \includegraphics[width=\linewidth]{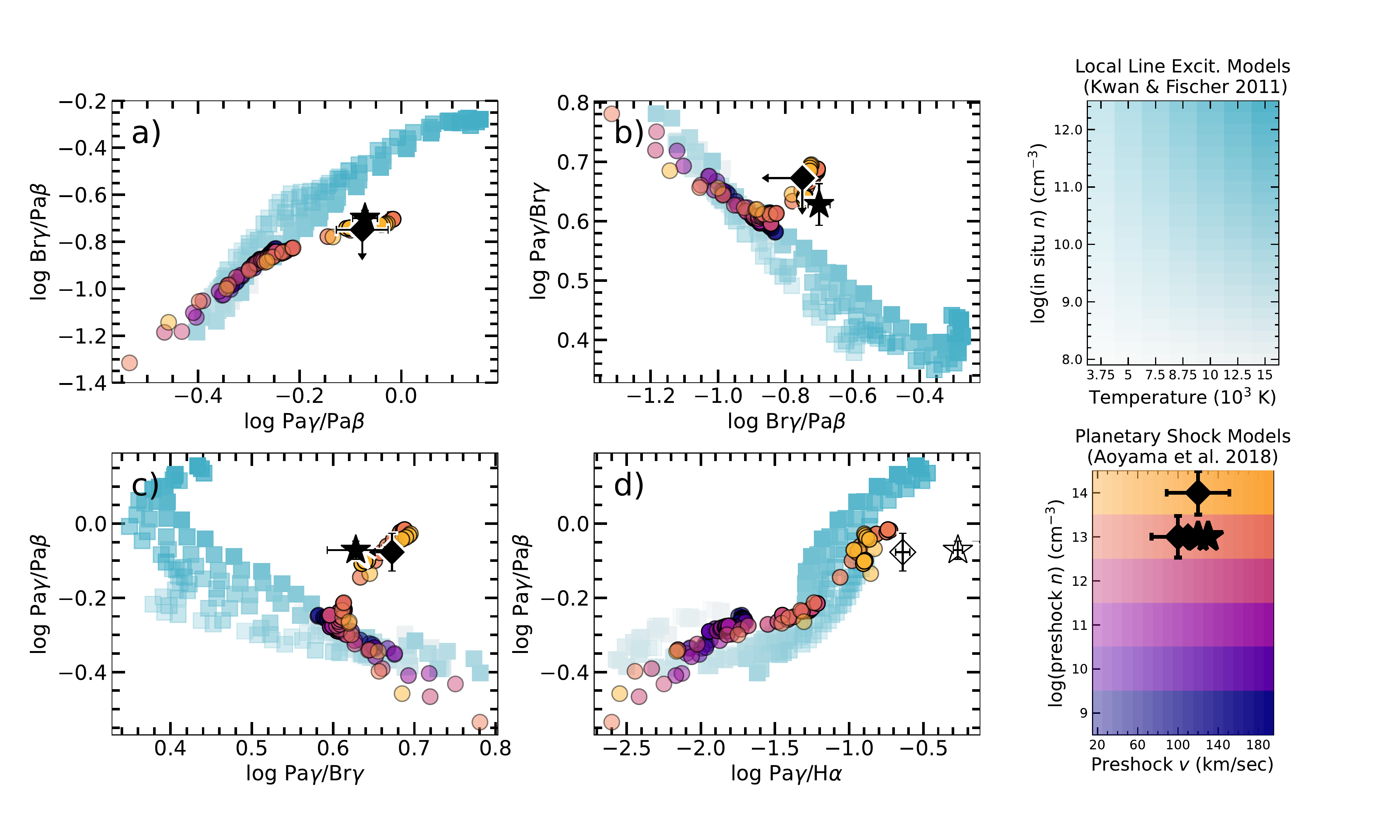}
    \caption{Predicted line ratios for local line excitation \citep[``stellar",][blue squares]{Kwan2011}, and accretion shock emission \citep[``planetary",][purple/yellow circles]{Aoyama2018} models. Delorme 1 (AB)b's NIR line ratios (panels a-c) are shown in black with markers (stars, diamonds) indicating epoch as in Figure~\ref{fig2}. Panel d shows Pa$\gamma$/H$\alpha$ line ratios, with Delorme 1 (AB)b values (unfilled markers) derived from our Pa$\gamma$ observations and the non-contemporaneous H$\alpha$ measurement of \citet{Eriksson2020}.   }
    \label{fig3}
\end{figure*}

\section{Discussion}
We have presented mass accretion rate estimates for Delorme 1 (AB)b derived from NIR hydrogen emission lines under two assumed scalings of \Lline to $\Lacc/\dot{M}$.  Accretion rate estimates for individual NIR lines agree with one another within the ``planetary'' and ``stellar'' accretion paradigms, with the exception of the ``stellar'' Pa$\beta$ accretion rate, which is marginally inconsistent with the other ``stellar" accretion estimates.      

In Figure~\ref{fig2}, we compare our NIR observations (diamonds/stars) with the marginally-resolved H$\alpha$ observations of \citet{Eriksson2020} (gray/black circles) and convert each to \Mdot using both ``stellar" (unfilled symbols, \citealt{Alcala2017}; dark gray, \citealt{Natta2004}) and ``planetary" (filled symbols, \citealt{Aoyama2019}; light gray, \citealt{Thanathibodee2019}) scaling relations. H$\alpha$ can originate from chromospheric activity, complicating its interpretation. \citet{Eriksson2020} found that the contribution to the H$\alpha$ line profile due to chromospheric activity should be minimal at this age, pointing toward Delorme 1 (AB)b experiencing ongoing accretion.
We find that our NIR \Mdot's generally agree with the \citet{Eriksson2020} estimates within uncertainties, albeit with slightly higher \Mdot values relative to the Balmer series, though our \PaB measurement is marginally inconsistent at the 1$\sigma$ level. 
Given the strength of the companion's NIR EWs relative to diagnostics measured for active low-mass stars \citep[$\sim0.04-0.05$ \AA; e.g.,][]{Schofer2019}, our results are most consistent with the presence of PMC accretion. 

\begin{figure*}[htp]
    \centering
    \includegraphics[width=\linewidth]{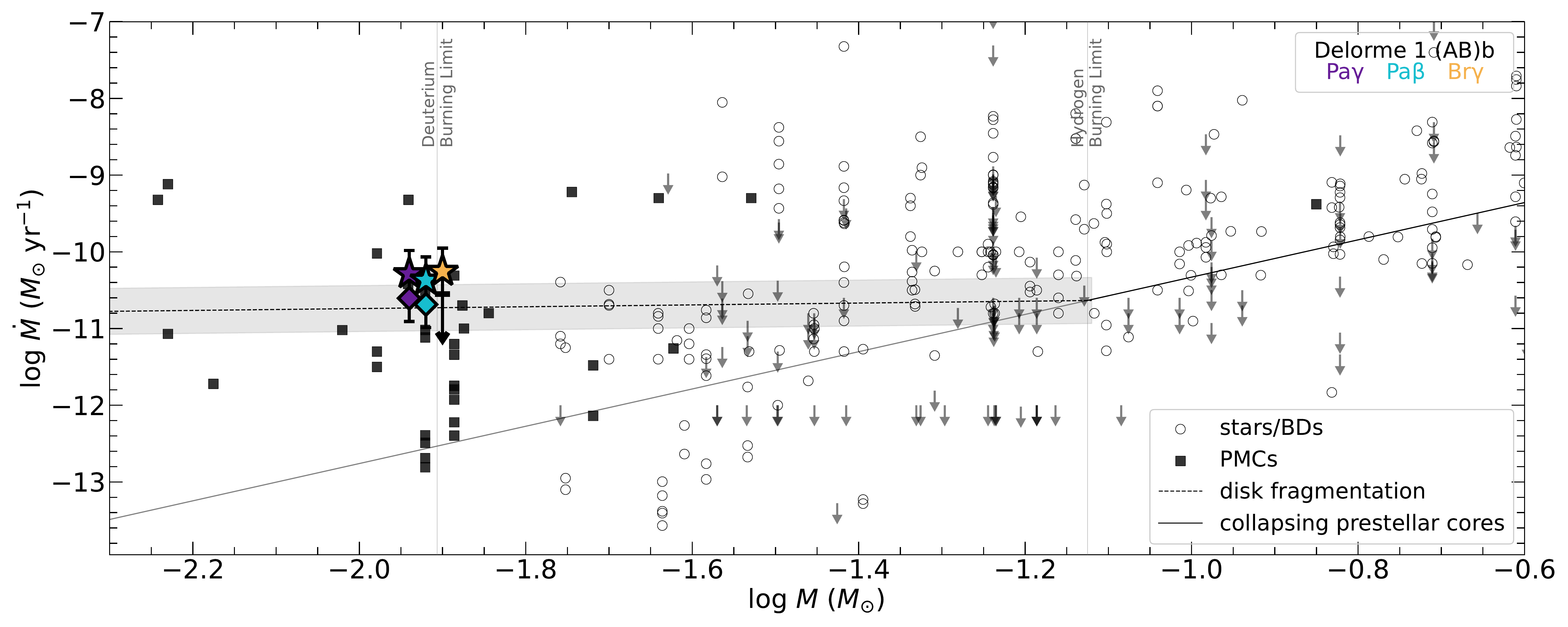}
    \caption{Mass accretion rate vs mass for all known isolated substellar accretors (gray), planetary mass companions (black squares), and a representative sample of low mass stars. Delorme 1 (AB)b's NIR derived \Mdot is highlighted (colored markers, symbol as in Fig~\ref{fig2}).  The canonical $\dot{M} \propto M^{2.1}$ \citep{Muzerolle2005} relation for higher mass objects (consistent with formation via collapsing prestellar cores) is shown (solid line), and a predicted relation for substellar objects formed via disk fragmentation  \citep[dashed line, $\alpha\sim0.001$, ][]{Stamatellos2015}. Delorme 1 AB(b) Pa$\gamma$ and Br$\gamma$ measurements have been offset in mass for clarity.}
    \label{fig4}
\end{figure*}

We find agreement between $\Mdotpla$ and $\dot{M}$(H$\alpha$ 10\%); both are $\sim$1.5 mag higher than $\dot{M}_\mathrm{ste}$.
As $\dot{M}$(H$\alpha$ 10\%) does not rely on scaling relationships, accurate continuum subtraction, or extinction, it is considered a robust independent measure of accretion \citep{White2003, Stelzer2007}, including for the lowest mass accreting protoplanets \citep[e.g., PDS 70b;][]{Haffert2019}. As noted by \citet{Alcala2014}, the empirical relationship between H$\alpha$ 10\% width and \Mdot \citep{Natta2004} has considerable scatter, and line luminosities should also be used when possible.  
However, the strong agreement between $\dot{M}$(H$\alpha$ 10\%) and $\Mdotpla$ could indicate that $\Mdotpla$ is a more accurate estimate of \Mdot for Delorme 1 (AB)b.
The marginal inconsistency in $\dot{M}_\mathrm{ste}$ could be a result of applying stellar scaling relations to an object accreting under a different paradigm; this is not seen in the $\Mdotpla$s. 

To independently determine the accretion paradigm most consistent with Delorme 1 (AB)b without a reliance on scaling relations, line ratios can be used.
NIR hydrogen lines are ideal for measuring accretion line ratios \citep[see][]{Edwards2013, Bary2008} due to their small line opacities, resulting in little blue or redshifted absorption from winds or infalling gas \citep{Folha2001, Edwards2006}.  By comparing observed line ratios to accretion model prediction, we can probe physical conditions of the emitting region such as number density, temperature, and infall velocity. Line ratios have discriminating power between different physical line emission sources, as different accretion models predict different line ratios. 
To this end, we consider two models: the local line excitation model of \citet{Kwan2011} and the planetary shock model of \citet{Aoyama2018}. As shown in Figure~\ref{fig3}, the predicted line ratios of post-shock gas in a planetary atmosphere \citep[planetary paradigm,][purple/yellow circles]{Aoyama2018} can vary from those predicted by local line excitation models developed to describe infalling accretion columns of T Tauri stars \citep[stellar paradigm,][green/blue squares]{Kwan2011}, allowing us to infer which model better matches observations, though there is some overlap for lower densities, where we are not able to distinguish between accretion paradigms. 
We calculate line ratios for each line pair and epoch (star/diamond symbols) over the whole emission range\footnote{In T Tauri stars, winds and outflow absorption can affect line ratios.  As such, residual line profiles selected over regions with no opacity effects are used to calculate line ratios \citep[see][]{Edwards2013}.  However, these are assumed to be negligible in PMCs.}.
In panel d, we include ratios with respect to published H$\alpha$ emission for context, noting these observations were not obtained contemporaneously with our NIR data. Line ratios may be affected by intrinsic and instrumental variability; therefore, inconsistency of the H$\alpha$ ratio with either model grid may not be indicative of variability in the physical conditions of the emitting region.

For all measurements, observed line ratios fall nearest the \citet{Aoyama2018} models and consistently diverge from the \citet{Kwan2011} models, suggesting that planetary scaling relations are likely more appropriate in this situation.  Therefore, we use the \citet{Aoyama2021} models and relations for further analysis.  
We extract all model physical input parameters consistent with observed line ratios within uncertainties.  We find that the best-fitting models have preshock velocities of $70-170$ km/s and number densities of $10^{13-14}$ cm$^{-3}$. While the preshock velocity is consistent with measured \Mdot's and assumed mass \citep[and radius; see Figure 13 of][]{Aoyama2020}, the number density is higher than expected for the measured \Mdot assuming a pure planetary shock model. This could be explained by shock emission with a low filling factor resulting from a magnetospheric accretion flow,  absorption in the post-shock region \citep{Hashimoto2020}, strong accretion column extinction \citep[][though they found that the \Mdot is too low for absorption by either gas or dust in the accretion flow]{Marleau2022}, or circumplanetary disk extinction in the line of sight \citep{Aoyama2020}.  High resolution (R$\sim$10,000) spectra will help disentangle the accretion flow geometry and shed light on the nature of the accretion shock, as resolved line profiles can distinguish between geometries \citep{Aoyama2020, Marleau2022}.

In Figure~\ref{fig4}, we show the \Mdot--$M$ relation for all known accreting substellar objects, together with low mass stars (Betti et al., in prep.).  The $\Mdotpla$'s for Delorme 1 (AB)b lie above the canonical $\Mdot\sim M^{2.1}$ \citep{Muzerolle2005} T Tauri star relation consistent with formation via collapsing prestellar cores.  The mass accretion rates are similar to other bound planetary mass companions (black squares),
whose previous accretion rate estimations mostly come from H$\alpha$ line luminosity or H$\alpha$ 10\% width. The location of these bound PMCs in \Mdot--$M$ space is consistent with model predictions of PMC formation through disk fragmentation in disks with low viscosities \citep{Stamatellos2015}. These models predict higher accretion rates; companions that form in dynamically unstable systems have larger than expected gas mass reservoirs, allowing them to accrete more material \citep{Stamatellos2015} for longer. The high \Mdot observed for Delorme 1 (AB)b suggests that it may have formed via disk fragmentation.  Its \Mdot is most consistent with \citet{Stamatellos2015} models with low disk viscosity ($\alpha \sim$0.001), and is comparable to PMCs with similar masses such as GSC~06214-00210~b, GQ~Lup~b, and DH~Tau~b, all of which have been theorized to have formed via disk fragmentation \citep{Stamatellos2015, Zhou2014}.

In summary, the strong Pa$\beta$, Pa$\gamma$, and Br$\gamma$ emission seen from  Delorme 1 (AB)b indicates strong ongoing mass accretion onto the PMC. 
Utilizing line ratios, we find that the NIR hydrogen emission is most consistent with models of planetary shock accretion, though the high predicted number density does not exclude magnetospheric accretion from occurring as well on the planetary surface. We conclude that higher \Mdot estimates derived from planetary scaling relations are more likely to reflect the true accretion rate, and the position of Delorme 1 (AB)b in \Mdot--$M$ space is consistent with formation via disk fragmentation.  This would account for its high accretion rate, which is consistent with low disk viscosity, likely resulting in slower disk evolution and perhaps explaining why this 30--45~Myr object is still actively accreting \citep[potentially a ``Peter Pan disk";][]{Silverberg2020}.  Detailed modeling of the planetary surface and disk will provide a clearer understanding of Delorme 1 (AB)b, and future observations of a wider range of line ratios will help constrain the nature of the accretion shock.  Forthcoming work (Betti et al, in prep) will present detections of NIR accretion for a comprehensive sample of accreting BDs and PMCs as well as observational \Lacc--\Lline empirical relationships for the substellar regime in order to help constrain substellar formation mechanisms.   
Delorme 1 (AB)b is a benchmark accreting PMC, with current observations and theoretical models suggesting the nature of its emission is in the planetary regime.

\acknowledgments
We thank the anonymous referee for their careful review.
We thank the SOAR support scientist, CTIO scientist Sean Points, for allocating some engineering time for our observations.
S.K.B. and K.B.F. acknowledge support from NSF AST-2009816.
G-DM acknowledges the support of the DFG priority program SPP 1992 ``Exploring the Diversity of Extrasolar Planets'' (MA~9185/1) and from the Swiss National Science Foundation under grant
200021\_204847 ``PlanetsInTime''.
Parts of this work have been carried out within the framework of the NCCR PlanetS supported by the Swiss National Science Foundation.

\vspace{5mm}
\facilities{SOAR(TripleSpec4.1)}

\software{astropy \citep{2013A&A...558A..33A,2018AJ....156..123A}, specutils \citep{specutils}, Spextool \citep{Vacca2003, Cushing2004}, Matplotlib \citep{Hunter2007}}

\bibliography{ref}
\bibliographystyle{aasjournal}

\end{document}